\documentclass[twocolumn,pre,floatfix]{revtex4}
\usepackage{psfrag,epsfig,amsfonts,amssymb,amsmath,wasysym,bm}
\usepackage{dcolumn}
\usepackage{bbold}
\usepackage[normalem]{ulem}
\usepackage{color}
\usepackage{tabularx}

\usepackage{enumitem} % enumerated lists

\usepackage{hyperref}
\newcommand{\ket}[1]{\lvert #1 \rangle} 	% ket
\newcommand{\bra}[1]{\langle #1 \rvert}	% bra
\newcommand{\ketP}[1]{\lvert #1 \rangle^{\! \prime}} 	% ket
\newcommand{\braP}[1]{{^\prime\!}\langle #1 \rvert}	% bra
\newcommand{\mc}{\mathcal}	% calligraphic
\newcommand{\const}{\operatorname{const}}	% constant

\newcommand{\<}{\left\langle}	% angular brackets
\renewcommand{\>}{\right\rangle}	% "

\newcommand{\e}{\mathrm{e}}		% Euler number
\newcommand{\I}{\mathrm{i}}		% imaginary unit
\newcommand{\rhoMC}{\rho_{\mathrm{mc}}}
\newcommand{\rhoeq}{\rho_{\mathrm{eq}}}
\newcommand{\rhoForward}{\rho_{\mathrm{f}}}

\newcommand{\rhoBackward}{\rho_{\mathrm{b}}}

\newcommand{\rhoTarget}{\rho_{\mathrm{T}}}
\newcommand{\rhoTargetPert}{\rho^\prime_{\mathrm{T}}}
\newcommand{\rhoRev}{\rho_{\mathrm{R}}}

\newcommand{\ldos}{u}

\newcommand{\dA}{\mathcal{A}}

\newcommand{\tr}{\mbox{Tr}}

\providecommand{\norm}[1]{\|#1\|}

\providecommand{\av}[1]{\mathbb{E}[#1]}
\providecommand{\avv}[1]{\mathbb{E}\!\left[#1\right]}

\renewcommand{\d}{\mathrm{d}} % differential
\newcommand{\lmat}{\left( \begin{matrix}}	% matrix begin
\newcommand{\rmat}{\end{matrix} \right)}	% matrix end
\definecolor{revcolor}{rgb}{0.1,0.4,0.8}

\begin{document}

\title{Predicting imperfect echo dynamics in many-body quantum systems}
\author{Lennart Dabelow}
%\email{ldabelow@physik.uni-bielefeld.de}
\author{Peter Reimann}
%\email{reimann@physik.uni-bielefeld.de}
\affiliation{Fakult\"at f\"ur Physik, 
Universit\"at Bielefeld, 
33615 Bielefeld, Germany}
\date{\today}

\begin{abstract}
Echo protocols provide a means to investigate the arrow of time in macroscopic processes.
Starting from a nonequilibrium state, the many-body quantum system under 
study is evolved for a certain period of time $\tau$.
Thereafter, an (effective) time reversal is performed that would -- if implemented 
perfectly -- take the system back to the initial state after another time period $\tau$.
Typical examples are nuclear magnetic resonance imaging and polarization 
echo experiments.
The presence of small, uncontrolled inaccuracies during the backward 
propagation results in deviations of the ``echo signal'' from the original evolution, 
and can be exploited to quantify the instability of nonequilibrium states and 
the irreversibility of the dynamics.
We derive an analytic prediction for the typical dependence of this echo signal 
for macroscopic observables on the magnitude of the inaccuracies and on
the duration $\tau$ of the process, and verify it in numerical examples.
\end{abstract}

\maketitle

\section{Introduction}
\label{sec:Intro}

Explaining the irreversibility of processes in macroscopic systems based on 
the time-reversible laws governing their microscopic constituents is a major 
task of statistical mechanics, dating all the way back to Boltzmann's 
$H$-theorem \cite{bol72} and Loschmidt's paradox \cite{los76}.
Besides its ontological dimension, this question is also intimately related to 
the special role of nonequilibrium states and their apparent instability in many-body 
systems, which generically tend towards equilibrium as time progresses.
Characterizing this instability within the realm of quantum mechanics 
is one goal of the present work.
Whereas a direct investigation of the pertinent states 
(i.e., Hilbert space vectors or density 
operators) can in principle have academic 
value, similarly to an analysis of phase-space 
points in classical systems, we focus here on \emph{observable} 
quantities that can be extracted 
from %the dynamics of 
macroscopic measurements.

Recently, so-called out-of-time-order correlators (OTOCs) have gained considerable 
attention as a suggestion to generalize concepts from classical chaos theory, notably 
Lyapunov exponents, to quantum systems \cite{mal16, swi18},
but the analogy is far from complete \cite{hum19, pil20}.
With this in mind, we follow an even simpler route here and consider so-called echo dynamics 
\cite{hah50}, where a quantum many-body system with Hamiltonian $H$ starts 
from a (pure or mixed) ``target state'' $\rhoTarget$ and is evolved 
forward in time for a certain period $\tau$ to reach the ``return state'' $\rhoRev$.
At this point, one switches to the inverted Hamiltonian $-H$, so that the 
direction of time is effectively reversed and the system evolves back 
towards the target state after another time period $\tau$.
By introducing inaccuracies (experimentally unavoidable imprecisions) 
during this backward evolution via a Hamiltonian of the form $H' := -H + \epsilon V$,
the system will not reach the original initial state $\rhoTarget$ again, but instead 
approach a perturbed state $\rhoTargetPert$ that deviates from it to some extent.
In summary,
\begin{equation}
\label{eq:ITR:Protocol}
	\rhoTarget \xrightarrow[H]{\;\;\;\;\tau\;\;\;\;} \rhoRev \xrightarrow[-H + \epsilon V]{\;\;\;\;\tau\;\;\;\;} \rhoTargetPert \,.
\end{equation}
The deviations between the initial and final states and their dependence on the 
time span $\tau$ and on the amplitude $\epsilon$ of the considered imperfections 
then quantify the instability of the nonequilibrium state and the irreversibility 
of the dynamics.

In case of a pure state, a seemingly direct probe is the Loschmidt echo, 
i.e., the overlap between the original state $\rhoTarget$ and the distorted 
echo state $\rhoTargetPert$ \cite{per84, gor06}.
(In case of a mixed state, the corresponding probe is the quantum 
fidelity.)
However, this quantity is practically inaccessible in a many-body system,
and, as emphasized above, we will instead concentrate on echoes of 
macroscopic observables in the following and compare their expectation 
values at the beginning and at the end of the proposed protocol 
\cite{gor06, fin14, sch16, sch18}.
Since macroscopic observables cannot distinguish between equivalent 
microstates, one inevitably has to operate in the nonequilibrium 
regime to be able to characterize the sensitivity of a many-body 
system towards imperfections.

Taken literally, the suggested protocol~(\ref{eq:ITR:Protocol})
can only have the status of a gedankenexperiment 
since we cannot practically revert the direction of time in a concrete dynamical setup,
and also an exchange of the Hamiltonian $H$ for $-H$ is still unphysical 
in many situations.
For example, for a gas of particles in a box, it would require negative particle masses.
Nevertheless, it is well known that an effective sign change of the Hamiltonian can be 
achieved in spin systems, which provides the mechanism underlying spin-echo 
and nuclear magnetic resonance (NMR) experiments \cite{hah50}.
In fact, the presence of imperfections, i.e., ``non-reversed components'' 
of the Hamiltonian, forms the basis of magnetic resonance imaging (MRI)
by exploiting that different imperfections in different tissues lead to 
distinct imperfect echo signals \cite{vla03}.
By means of so-called magic- or polarization-echo techniques, these ideas 
have also been extended to interacting spin systems 
\cite{sch69, rhi70, rhi71, zha92, kim92, haf96, lev98, usa98}, 
employing suitably adapted radio-frequency external fields during 
the ``backward'' phase of the evolution.
More generally, pulse sequences and time-dependent forces have been used to scan quite notable parameter ranges of (effective) spin Hamiltonians in a variety of experimental setups \cite{gar17, wei18, wei19, nik20}.
Yet another
experimental approach towards an effective time 
reversal consists in tuning a cold atomic gas across a 
Feshbach resonance \cite{wid08, cuc10, wei13}.
Given the neglected unreversed corrections as well as the 
sophisticated experimental setups necessary, some sort of 
imperfections of the form considered here are clearly 
unavoidable and sometimes, like in MRI, even desired.

Furthermore, an alternative view on the suggested protocol may be as follows:
For a many-body system with Hamiltonian $\tilde H$ ($= -H$ in the language 
from above), suppose that we are given an initial state $\rhoRev$ (previously 
the return state), for which it is known that the state obtained after time $\tau$, 
$\rhoTarget = \e^{-\I \tilde H \tau} \rhoRev \e^{\I \tilde H \tau}$ $(\hbar = 1)$, 
is out of equilibrium.
Comparing the time evolution from $\rhoRev$ under the Hamiltonian 
$\tilde H$ with the dynamics obtained from a perturbed Hamiltonian 
$\tilde H' = \tilde H + \epsilon V$, we achieve the same effect as in 
the above echo gedankenexperiment~(\ref{eq:ITR:Protocol}).
However, there is no need for a ``backward Hamiltonian'' or some other 
sort of (experimentally difficult) reversal procedure.
From a physical point of view, it is thus irrelevant whether the return 
state $\rhoRev$ was obtained by an explicit unitary time evolution or 
some other preparation method.
Its only relevant property is that it reaches a manifestly out-of-equilibrium 
state within an accessible time scale $\tau$ in order for the effects of 
imperfections to become macroscopically visible.
This setting is silently included in the following, even though we will 
employ the language of the echo protocol (\ref{eq:ITR:Protocol})
in the remainder of this work.

In the next Sec.~\ref{sec:Setting}, we set the stage and introduce 
the suggested echo protocol~(\ref{eq:ITR:Protocol}) and the considered 
imperfections $V$ in detail.
Our main result, an analytical prediction characterizing the decay of 
echo signals under generic time-reversal inaccuracies, is derived 
in Sec.~\ref{sec:Results}.
Thereafter, we verify this result numerically in an explicit spin chain 
model in Sec.~\ref{sec:Example}.
In Sec.~\ref{sec:Conclusions}, we conclude by summarizing the ideas 
and relating them to numerical and experimental results from the literature.

\section{Setup}
\label{sec:Setting}

To begin with, we represent the Hamiltonian $H$ appearing in 
(\ref{eq:ITR:Protocol}) in terms of its eigenvalues and eigenvectors as
\begin{equation}
\label{eq:H}
	H = \sum_n E_n \, \ket{n} \bra{n} \,.
\end{equation}
Given some initial state $\rho(0)$ at time $t = 0$,
the state at any later time $t>0$ follows as $\rho(t) := \e^{-\I H t} \rho(0) \e^{\I H t}$.
Being interested in some macroscopic observable (see introduction) in the form
of a self-adjoint operator $A$, and denoting its expectation value 
in an arbitrary state $\rho$ by $\langle A \rangle_{\!\rho} := \tr[ \rho A ]$,
the actual time evolution can thus be written as
\begin{equation}
\label{eq:TimeEvo:A}
	\< A \>_{\!\rho(t)} = \sum_{m, n} \e^{\I (E_n - E_m) t} \, \bra{m} \rho(0) \ket{n} \, \bra{n} A \ket{m} \,.
\end{equation}
Despite the quasiperiodic nature of the right-hand side, a many-body system will 
usually equilibrate \cite{rei08, lin09, sho12} and spend most of the time close to 
the time-averaged state $\rhoeq$ with
\begin{equation}
	\bra{m} \rhoeq \ket{n} := \delta_{mn} \bra{n} \rho(0) \ket{n} \,,
\end{equation}
which we refer to as the equilibrium state in the following.

In order to study the effect of imperfections in terms of expectation values of the 
observable $A$, the considerations from Sec.~\ref{sec:Intro} imply that the system 
must spend some time away from equilibrium, i.e., there must be a reasonable 
time interval during which $\<A\>_{\!\rho(t)}$ differs distinctly from $\< A \>_{\!\rhoeq}$.
Therefore, we focus on these deviations from equilibrium, denoted by the symbol
\begin{equation}
\label{eq:DevAFromEq}
	\dA(t) := \<A\>_{\!\rho(t)}  - \< A \>_{\!\rhoeq} .
\end{equation}
Provided that the system is out of equilibrium at time $t = 0$, we then ask how 
special this situation is by investigating how hard it is to return to this state by 
an effective, but possibly imperfect reversal of time after the system has 
relaxed for a certain period $\tau$ as detailed in the 
protocol~\eqref{eq:ITR:Protocol}.

In the absence of any imperfections ($\epsilon = 0$), the system traces out 
the same trajectory in the forward and backward stages, such that
\begin{equation}
\label{eq:PTR}
	\dA(\tau + t) = \dA(\tau - t)
\end{equation}
for $t \in [0, \tau]$,
which constitutes our reference dynamics.
It is reasonable to expect that
uncontrolled inaccuracies in the time-reversed dynamics will
generically push the system closer to equilibrium because they spoil 
the fine-tuned correlations between state and observable needed for 
nonequilibrium conditions.
Hence, the backward dynamics will usually lie closer to the equilibrium 
state than the forward one,
\begin{equation}
\label{eq:ITR}
	\lvert \dA(\tau + t) \rvert \lesssim \lvert \dA(\tau - t) \rvert
	\qquad
	(0 \leq t \leq \tau) \,.
\end{equation}
The sensitivity of the deviations between the perfect and 
perturbed dynamics 
with respect to the magnitude $\epsilon$ of the inaccuracies
is thus an indicator for the 
chaoticity and irreversibility of the many-body dynamics.
The faster $\dA(\tau+t)$ decays with $\epsilon$ compared to $\dA(\tau - t)$, 
the harder it is to design a reversible process and the more extraordinary or special 
are the nonequilibrium states.
Consequently, the relative echo signal $\dA(\tau + t) / \dA(\tau - t)$ for times $t \in [0, \tau]$ will 
be our principal object of study in this work, most importantly in the region around 
the revival or echo peak at $t \approx \tau$, where deviations from equilibrium 
will be most pronounced.

We denote the time-dependent state of the system in the forward and backward phases by
\begin{subequations}
\label{eq:ITR:Phases}
\begin{alignat}{2}
\label{eq:ITR:Phases:Forward}
	\rhoForward(t) &:=& \,\,\e^{-\I H t} &\,\rhoTarget\, \e^{\I H t} \,, \\
\label{eq:ITR:Phases:Backward}
	\rhoBackward(t) &:=& \,\,\e^{\I (H-\epsilon V) t} &\,\rhoRev\, \e^{-\I (H-\epsilon V) t} \,,
\end{alignat}
\end{subequations}
respectively.
We also write $\rho(t)$ to refer to the state during the entire process, i.e.,
$\rho(t) := \rhoForward(t)$ for $t \in [0, \tau]$ and $\rho(t) := \rhoBackward(t - \tau)$ for $t \in [\tau, 2\tau]$.

An implicit assumption in all what follows is 
that the considered many-body 
system is finite and exhibits a well-defined 
macroscopic energy $E$.
Consequently, the state $\rho(t)$ at any time 
can only significantly populate 
energy levels within a macroscopically 
small energy window 
\begin{eqnarray}
I_E := [E - \Delta E, E]
\label{x3}
\end{eqnarray}
and the imperfections are assumed to be 
sufficiently small so that they do not 
modify this window.
In addition, it is taken for granted that 
the density of states (DOS) of $H$ is 
approximately constant throughout this 
energy window, $D_0 \approx \const$,
and that the same holds for the (negative) 
imperfect backward Hamiltonian 
$H - \epsilon V$ with eigenvalues 
$E^\prime_\nu$ and eigenstates 
$\ketP{\nu}$. 

Focusing on the dynamics during the backward phase
from (\ref{eq:ITR:Phases:Backward}), we can use the 
transformation matrix $U_{\mu k} := \braP{\mu} k \rangle$ 
between the eigenbases of the forward and backward Hamiltonians
to write the time-dependent expectation values of the 
observable $A$ -- similarly as in (\ref{eq:TimeEvo:A}) --
as
\begin{align}
	\< A \>_{\!\rhoBackward(t)}
		&= \sum_{\mu, \nu} \braP{\mu} \rhoBackward(t) \ketP{\nu} \, \braP{\nu} A \ketP{\mu} \\
		&= \sum_{\mu, \nu} \sum_{k,l,m,n} \e^{-\I(E'_\nu - E'_\mu)t} \e^{\I(E_l - E_k) \tau} \notag \\
		& \qquad \times \bra{k} \rhoTarget \ket{l} \, \bra{m} A \ket{n} \, 
			U_{\mu k} \, U^{*}_{\nu l} \, U_{\nu m} \, U^{*}_{\mu n} \,.
\end{align}
Employing the assumed constant DOS for $H$ and $H-\epsilon V$, 
we can approximately identify energy differences 
$E^\prime_\nu - E^\prime_\mu \simeq E_\nu - E_\mu$ 
of the two Hamiltonians within in the relevant time scales \cite{rei19, rei19a}, so that
\begin{equation}
\label{eq:ITR:TimeEvoBack:A}
\begin{aligned}
	\< A \>_{\!\rhoBackward(t)}
		&= \sum_{\mu, \nu} \sum_{k,l,m,n} \e^{-\I(E_\nu - E_\mu)t} \e^{\I(E_l - E_k) \tau} \\
		& \qquad \times \bra{k} \rhoTarget \ket{l} \, \bra{m} A \ket{n} \, 
			U_{\mu k} \, U^{*}_{\nu l} \, U_{\nu m} \, U^{*}_{\mu n} \,.
\end{aligned}
\end{equation}

We recall that the Hamiltonian $H$ corresponds to a given many-body quantum 
system, whereas the perturbation $V$ describes 
uncontrolled and/or unknown inaccuracies in the time-reversal procedure.
In this spirit, we thus model our ignorance about these imperfections by
an ensemble of random operators $V$, such that the matrix 
elements $\bra{m} V \ket{n}$ of $V$ in the eigenbasis of $H$ 
become random variables.
The actually considered $V$ ensembles are inspired by the 
structure of typical perturbations, featuring possible sparsity as well 
as an interaction strength depending on the energy difference 
between the coupled states (``bandedness'') 
\cite{gen12, beu15, kon15, bor16, jan19}.
Requiring Hermiticity, $\bra{m} V \ket{n} = \bra{n} V \ket{m}^*$, 
and assuming independence of the $\bra{m} V \ket{n}$ for $m \leq n$, 
this suggests the general form
\begin{equation}
\label{eq:ITR:PDFV}
	\d P_{mn}(v) := \d \mu_{\lvert E_m - E_n \rvert}(v)
\end{equation}
for the probability measures of the $\bra{m} V \ket{n}$'s with $m < n$.
Here, $\{ \d\mu_\Delta \}_{\Delta > 0}$ denotes a family of probability 
measures on $\mathbb{R}$ or $\mathbb{C}$ with mean zero and 
variance $\sigma_v^2(\Delta)$, so that the smooth function 
$\sigma_v^2(\Delta)$ 
captures the announced bandedness of the 
interaction matrix.
Likewise, for $m=n$ the probability measure $\d P_{nn}(v) := \d\mu_0(v)$ 
of the (real-valued) diagonal elements $\bra{n} V \ket{n}$ is assumed 
to have vanishing mean (otherwise the perturbation would induce 
an energy shift) and finite variance.

To obtain a useful prediction regarding the behavior of an actual system, 
we first compute the average effect of such a perturbation.
In a second step, we establish that the resulting prediction satisfies a 
concentration of measure property, meaning that in a sufficiently 
high-dimensional Hilbert space, a particular realization of the ensemble 
becomes practically indistinguishable from the average behavior.
More precisely, deviations from the average will turn out to be suppressed 
in the number $N_v$ of eigenstates of $H$ that get mixed up by the 
perturbation $\epsilon V$, to be defined explicitly in Eq.~\eqref{eq:AvgU2:FWHM} below.
Due to the extremely high level density in generic many-body systems, 
this number $N_v$ is typically exponentially large in the system's 
degrees of freedom $f$ if the perturbation has 
any appreciable effect at all \cite{dabYY},
\begin{equation}
\label{eq:NumMixedLevels}
	N_v = 10^{\mathcal{O}(f)} \gg 1 \,.
\end{equation}

\section{Results}
\label{sec:Results}

According to Eq.~\eqref{eq:ITR:TimeEvoBack:A}, averaging the echo signal over all possible realizations of the $V$ ensemble requires an average over four transformation matrices $U_{\mu k}$, the overlap of the eigenvectors $\ket{k}$ of $H$ and $\ketP{\mu}$ of $H - \epsilon V$.
These overlaps inherit their distribution from the distribution~\eqref{eq:ITR:PDFV} of the $V$ matrix elements.
Writing $\av{ \,\cdots }$ for the average over all $V$'s, one finds
\begin{equation}
\label{eq:ITR:AvgU4}
\begin{aligned}
	& \av{ U_{\mu_1 k_1} U_{\mu_2 k_2} U^{*}_{\mu_1 l_1} U^{*}_{\mu_2 l_2} }
		= \delta_{k_1 l_1} \delta_{k_2 l_2} d^{\mu_1 \mu_2}_{k_1 k_2} \\
	& \qquad\qquad\qquad
			+ \delta_{k_1 l_2} \delta_{k_2 l_1} \left( \delta_{\mu_1 \mu_2} d^{\mu_1 \mu_2}_{k_1 k_2} + f^{\mu_1 \mu_2}_{k_1 k_2} \right)
\end{aligned}
\end{equation}
in the limit of sufficiently weak $\epsilon$ \cite{dabYY}.
Here
\begin{align}
\label{eq:ITR:AvgU4:d}
	d^{\mu\nu}_{k l} &:= \ldos(\alpha \epsilon^2, E_\mu -E_k) \, \ldos(\alpha \epsilon^2, E_\nu - E_l) \ ,
 \displaybreak[0] \\
\label{eq:ITR:AvgU4:f}
	f^{\mu\nu}_{k l} & := - \left( 
          \alpha \epsilon^2 / 2 \pi D_0 \right) \, 
           \ldos(\alpha \epsilon^2, E_\mu - E_k) \, \ldos(\alpha \epsilon^2, E_\nu - E_l) \\
		& \hspace{-20pt} \times\! \frac{4 \alpha^2 \epsilon^4 + (E_k \!-\! E_l)^2 + (E_\mu \!-\! E_\nu)^2 - (E_k \!+\! E_l \!-\! E_\mu \!-\! E_\nu)^2 }{ \left[ (E_\mu - E_l)^2 + \alpha^2 \epsilon^4 \right] \left[ (E_\nu - E_k)^2 + \alpha^2 \epsilon^4 \right] } , \notag
\end{align}
and $\alpha := \pi \bar \sigma_v^2 D_0$, where $\bar \sigma_v^2$ denotes the mean value of $\sigma_v^2(\Delta)$, introduced below~\eqref{eq:ITR:PDFV}, for small argument.
Furthermore, the function
$u(\alpha \epsilon^2, E_\mu - E_k) := \av{ \lvert U_{\mu k} \rvert^2 }$ represents
the second moment of the transformation 
matrices $U_{\mu k}$ and is given
by the Breit-Wigner distribution
\begin{equation}
\label{eq:ITR:AvgU2}
	\ldos(\gamma, E) = \frac{ \gamma }{ \pi D_0 ( \gamma^2 + E^2 ) } \,.
\end{equation}
Hence we can identify
\begin{equation}
\label{eq:AvgU2:FWHM}
	N_v := 2 \alpha D_0 \epsilon^2 = 2\pi \bar\sigma_v^2 D_0^2 \epsilon^2
\end{equation}
as the full width at half maximum of the average overlap $\avv{\lvert U_{\mu k}\rvert^2}$ between eigenvectors of $H$ and $H - \epsilon V$, quantifying the number of energy eigenstates mixed by the perturbation as introduced above Eq.~\eqref{eq:NumMixedLevels}.

Exploiting~\eqref{eq:ITR:AvgU4} in the average of~\eqref{eq:ITR:TimeEvoBack:A}, we obtain
\begin{equation}
\label{eq:ITR:AvgTimeEvo}
\begin{aligned}
	&\avv{ \< A \>_{\!\rhoBackward(t)} }
		= \sum_\mu d^{\mu\mu}_{kl} \bra{k} \rhoTarget \ket{k} \bra{l} A \ket{l} + \sum_{\mu,\nu} \e^{-\I (E_\nu - E_\mu) t} \times \\
		&\; \times \!\sum_{k,l} \!\left[ \e^{\I (E_l - E_k) \tau} d^{\mu\nu}_{kl} \bra{k} \rhoTarget \ket{l} \bra{l} A \ket{k}  + f^{\mu\nu}_{kl} \bra{k} \rhoTarget \ket{k} \bra{l} A \ket{l} \right]
\end{aligned}
\end{equation}
If we make use of the constant DOS once again
(see below (\ref{x3})),
which allows us to shift summation indices, 
we find that
\begin{align}
\label{x2}
	\sum_{\mu} d^{\mu\mu}_{kl}
		&= \ldos(2\alpha\epsilon^2, \omega_{kl}) \,, \\
	\sum_{\mu,\nu} \e^{-\I (E_\nu - E_\mu) t} d^{\mu\nu}_{kl}
		&= \e^{\I \omega_{kl} t} \e^{-2 \alpha \lvert t \rvert \epsilon^2} \,, \\
	\sum_{\mu,\nu} \e^{-\I (E_\nu - E_\mu) t} f^{\mu\nu}_{kl}
		&= - \ldos(2\alpha\epsilon^2, \omega_{kl}) \e^{ -2 \alpha \lvert t \rvert \epsilon^2 } \\
		& \quad \times \!\left[ \cos( \omega_{kl} t) + \tfrac{ 2 \alpha \epsilon^2 }{ \omega_{kl} } \sin( \omega_{kl} \lvert t \rvert) \right] \notag 
\end{align}
with $\omega_{kl} := E_k - E_l$.
Substituting into~\eqref{eq:ITR:AvgTimeEvo} yields
\begin{equation}
\label{eq:ITR:AvgTimeEvo2}
\begin{aligned}
	\avv{ \< A \>_{\!\rhoBackward(t)} } - \< A \>_{\!\tilde\rho}
		&=  \e^{-2 \alpha \lvert t \rvert \epsilon^2 } \left[ \< A \>_{\!\rhoForward(\tau - t)} - \< A \>_{\!\tilde\rho}\right]
  + R(|t|)
%		 \\
%		&= \< A \>_{\!\rhoForward(\tau - t)} \e^{-2 \alpha \lvert t \rvert \epsilon^2 } \\
%		& \quad + \< A \>_{\!\tilde\rho} \left( 1 - \e^{-2 \alpha \lvert t \rvert \epsilon^2 } \right) + R(t)
\end{aligned}
\end{equation}
with the locally averaged equilibrium state $\tilde\rho$ given by 
$\bra{m} \tilde\rho \ket{n} := \delta_{mn} \sum_k \ldos(2\alpha\epsilon^2, E_n - E_k) \bra{k} \rhoTarget \ket{k}$ 
and
\begin{equation}
\label{eq:ITR:AvgTimeEvo:Rest}
\begin{aligned}
	R(t)
		&:= \e^{-2 \alpha t \epsilon^2 }
\sum_{k,l} \bra{k} \rhoTarget \ket{k} \bra{l} A \ket{l} \, \ldos(2 \alpha \epsilon^2, \omega_{kl}) 
\\
&\qquad\quad \times \left\{ 1 - \cos(\omega_{kl} t) - \tfrac{ 2 \alpha \epsilon^2 }{ \omega_{kl} } \sin( \omega_{kl} t) \right\} .
\end{aligned}
\end{equation}
Note that $\tilde\rho$ is the state approached for 
large times and can thus be identified with the 
equilibrium state $\rhoeq$ from~\eqref{eq:DevAFromEq}.
It usually corresponds to the pertinent thermal 
state \cite{deu91,rei15,gog16,nat18}.
We also observe that $R(t)$ vanishes at 
$t=0$ and for $t\to\infty$.
Furthermore, as shown in Appendix~\ref{app:AvgTimeEvo:Rest:Bound},
its magnitude can be bounded from above 
for arbitrary $t\geq 0$ according to
\begin{equation}
\label{eq:AvgTimeEvo:Rest:Bound}
	R^2(t) 
\leq \frac{\delta^2\!A}{50 \, N_v}
\ ,
\end{equation}
where $N_v$ is the width of the eigenvector 
overlaps from~\eqref{eq:AvgU2:FWHM}, and
\begin{equation}
\label{eq:ETHViolation}
	\delta^2\! A := \sum_{n : E_n \in I_E} \left( A_{nn} - \< A \>_{\!\rhoMC} \right)^2
\,,
\end{equation}
where $\rhoMC$ denotes the microcanonical density operator corresponding to the energy window from~\eqref{x3}.
For generic (non-integrable) Hamiltonians $H$, essentially
all observables $A$ of actual interest are expected to satisfy
the so-called eigenstate thermalization hypothesis (ETH)
\cite{dal16}. Hence, the right-hand side of (\ref{eq:ETHViolation}) 
can be roughly estimated as $\norm{A}^2$ (with $\norm{A}$ 
denoting the operator norm of $A$, i.e., the largest eigenvalue in modulus)
and $R(t)$ in (\ref{eq:ITR:AvgTimeEvo2}) can be
neglected according to~{\eqref{eq:NumMixedLevels} and (\ref{eq:AvgTimeEvo:Rest:Bound}).

For integrable systems, the relevant observables are
still expected to satisfy the so-called weak ETH
\cite{bri10,ike13,alb15},  and thus the right hand side of
(\ref{eq:ETHViolation}) 
can be roughly estimated as $\norm{A}^2 N/f$,
where $N$ is the number of energy levels $E_n$
contained in $I_E$, while $f$ counts the degrees
of freedom of the considered system and therefore
scales as $\ln(N)$ (see also (\ref{eq:NumMixedLevels})).
Again, one can conclude from (\ref{eq:AvgTimeEvo:Rest:Bound}) 
that $R(t)$ in (\ref{eq:ITR:AvgTimeEvo2}) amounts in many 
cases to a small correction.

Finally, we point out that the bound 
(\ref{eq:AvgTimeEvo:Rest:Bound})
is still rather loose since the oscillating character 
of the summands in~\eqref{eq:ITR:AvgTimeEvo:Rest} 
with respect to both $\omega_{kl}$ and $t$ will usually 
result in very strong ``accidental cancellation'' effects,
which are entirely disregarded in our derivation
of the ``worst case'' bound (\ref{eq:AvgTimeEvo:Rest:Bound})
in Appendix~\ref{app:AvgTimeEvo:Rest:Bound}.

\begin{figure*}
\includegraphics[scale=1]{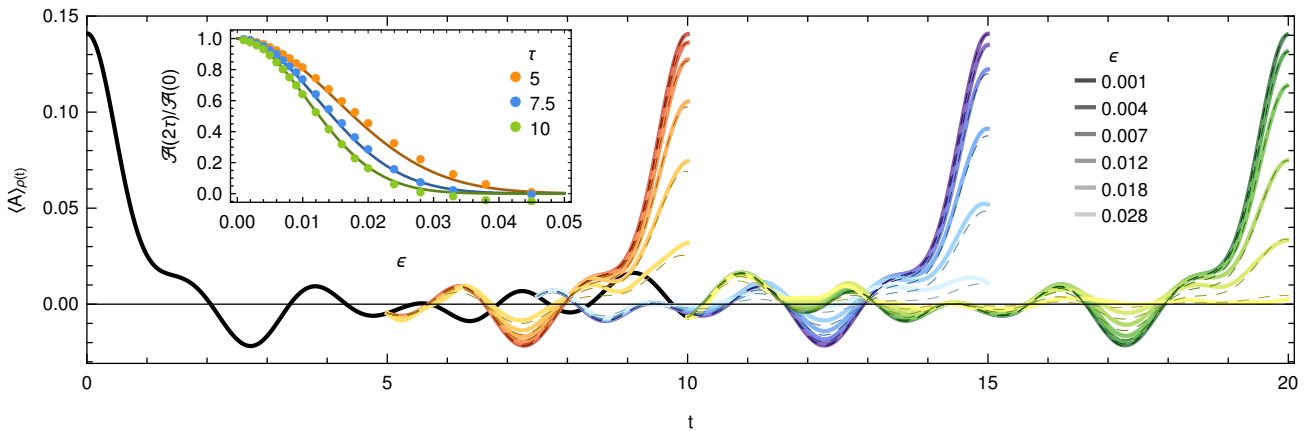}
\caption{
Time-dependent expectation values of the staggered magnetization (\ref{eq:IIC:Example:A})
for the spin-$\frac{1}{2}$ XXX chain (\ref{eq:IIC:Example:H}) with $L = 14$
under the ``imperfect reversal'' protocol~\eqref{eq:ITR:Protocol} with various 
perturbation strengths $\epsilon$ and reversal times $\tau$.
The initial condition is chosen as a filtered N\'{e}el target state 
$\rhoTarget = \ket{\psi}\bra{\psi}$ according to Eq.~\eqref{eq:ITR:Example:FilteredNeelState}
with $\sigma_{\psi}=1.3$.
The imperfection Hamiltonian $V$ in \eqref{eq:ITR:Protocol}
is of the ``spin-glass'' form~\eqref{eq:IIC:Example:V}.
Solid lines correspond to the numerical results using exact diagonalization.
Dashed lines show the prediction for the backward (echo) dynamics according
 to Eqs.~\eqref{eq:ITR:TypEcho} and (\ref{x4}).
Time reversal is initiated after time $\tau = 5$ (red-toned curves), 
$\tau = 7.5$ (blue-toned), or $\tau = 10$ (green-toned).
Inset: Ratio $\mc A(2\tau) / \mc A(0)$ of the echo peak height 
(at time $t=2\tau$) and the 
initial value (at $t=0$) as a function of the perturbation 
strength $\epsilon$ for the different reversal times $\tau$.
Data points are the numerical solutions, solid lines are the analytical 
prediction from~\eqref{eq:ITR:TypEcho}  and (\ref{x4}).
}
\label{fig:ITR:SpinES}
\end{figure*}

Indeed, we have not been able to identify any
specific example of practical interest where
the last term in (\ref{eq:ITR:AvgTimeEvo2})
plays a significant role.
Accordingly, this term is henceforth considered as
negligible.
With~\eqref{eq:DevAFromEq}, \eqref{eq:ITR:Phases}, 
and \eqref{eq:NumMixedLevels}
we thus arrive
at the first key result of this section,
\begin{equation}
\label{eq:ITR:AvgEcho}
	\frac{ \avv{ \dA(\tau + t) } }{ \dA(\tau - t) }
		= \e^{-2 \alpha t \epsilon^2 }
		\qquad
		(0 \leq t \leq \tau) \,,
\end{equation}
which quantifies the average effect of imprecisions 
during the backward evolution within the considered 
ensembles of $V$'s and for small enough $\epsilon$.
Analogously to Ref.~\cite{dabYY}, one can then proceed 
to derive a bound for the variance of $\dA(\tau + t)$.
This leads to
\begin{equation}
\label{eq:ITR:Variance}
	\avv{ \dA(\tau \!+\! t)^2 } - \avv{ \dA(\tau \!+\! t) }^{\,2}
		\leq \frac{C_v \,\norm{A}^2}{N_v} \,,
\end{equation}
where $C_v$ is a constant of order $10^3$ or less.
In view of~\eqref{eq:NumMixedLevels}, the 
variance~\eqref{eq:ITR:Variance} of $\dA(\tau + t)$ 
with $V$ is thus exponentially small in the number 
of degrees of freedom,
establishing a so-called concentration of measure
property of the considered ensembles of $V$ operators.
For instance, 
invoking Chebyshev's inequality from probability theory, 
the estimate~(\ref{eq:ITR:Variance})
implies that the 
probability for $\dA(\tau + t)$ to differ by more 
than $\norm{A} / N_v^{1/3}$ from the average 
$\av{ \dA(\tau + t) }$ at a certain instant in 
time $t$ is less than $C_v / N_v^{1/3}$.
This is our second key result of this section,
promoting Eq.~\eqref{eq:ITR:AvgEcho} 
from a mere 
statement about the ensemble average to a 
prediction for individual realizations:
Since deviations from the average are extremely 
rare for reasonably large many-body systems, we 
can conclude that
\begin{equation}
\label{eq:ITR:TypEcho}
	\frac{ \dA(\tau + t) }{ \dA(\tau - t) }
		= \e^{-2 \alpha t \epsilon^2 }
		\qquad
		(0 \leq t \leq \tau)
\end{equation}
is an excellent approximation for the vast majority 
of time-reversal inaccuracies $V$ captured by 
the considered ensembles.
This relation for the echo dynamics~\eqref{eq:ITR:Protocol} 
under an imperfect backward Hamiltonian constitutes 
our main result.
It asserts that the echo signal is exponentially 
suppressed in the propagation time $t$ and the 
intensity $\epsilon^2$ of the imperfections.

\section{Example}
\label{sec:Example}

We consider a spin-$\frac{1}{2}$ XXX chain model with Hamiltonian
\begin{equation}
\label{eq:IIC:Example:H}
	H = - \sum_{i = 1}^{L-1} \bm\sigma_i \cdot \bm\sigma_{i+1} \,,
\end{equation}
where $\bm\sigma_i = (\sigma^x_i, \sigma^y_i, \sigma^z_i)$ 
is a vector of Pauli matrices acting on site $i$.
For the perturbation $V$, we choose
\begin{equation}
\label{eq:IIC:Example:V}
	V = \sum_{i<j} \sum_{\alpha, \beta = 1}^3 J_{ij}^{\alpha\beta} \sigma_i^\alpha \sigma_j^\beta \,,
\end{equation}
where the couplings $J_{ij}^{\alpha\beta}$ are drawn independently from a normal 
distribution (unbiased Gaussian with unit variance).
As the observable, we take the staggered magnetization
\begin{equation}
\label{eq:IIC:Example:A}
	A = \frac{1}{L} \sum_{i=1}^L (-1)^i \, \sigma_i^z \,.
\end{equation}
Turning to the initial (target) state $\rhoTarget = \ket{\psi}\bra{\psi}$, 
let us first consider a N\'{e}el state of the form 
$\ket{\bar\psi} := \ket{ \downarrow \uparrow \downarrow \uparrow \cdots }$.
In order to account for the requirement that the density of states 
should be approximately constant (see below Eq. (\ref{x3})), 
we rescale the probability amplitudes 
$\langle n | \bar\psi \rangle$ according to the 
corresponding energy eigenvalues $E_n$ with a Gaussian 
weight of zero mean and standard 
deviation $\sigma_{\psi}$, resulting in 
\begin{equation}
\label{eq:ITR:Example:FilteredNeelState}
	\ket{\psi} := C \sum_n \e^{-E_n^2 / 2 \sigma_\psi^2} \, \langle n | \bar\psi \rangle \, \ket{n} \,,
\end{equation}
where the normalization constant $C$ is chosen such that $\langle \psi | \psi \rangle = 1$.
As discussed around Eq.~\eqref{x3}, the large isolated systems we have in mind (e.g., an MRI sample) are expected to exhibit a macroscopically well-defined energy, so that $\sigma_\psi$ should lie below the measurement resolution.
In particular, while it is possible to prepare a clean N\'{e}el state in few-body experiments with cold atoms, such attempts will most likely result in a filtered or coarse-grained variant as the degrees of freedom increase.

Quantitatively, for the example in  Fig.~\ref{fig:ITR:SpinES} we chose
a chain length of $L=14$ and a standard deviation of $\sigma_{\psi}=1.3$,
so that the state (\ref{eq:ITR:Example:FilteredNeelState})
is focused around energy $E = 0$ with about 
$15\,\%$ of the total $2^L = 16\,384$ levels $E_n$ within $\pm\sigma_\psi$.
This procedure reduces the staggered magnetization $A$ of $\ket{\psi}$ 
compared to $\ket{\bar\psi}$, but still gives an appreciably ouf-of-equilibrium 
expectation value (see also Fig.~\ref{fig:ITR:SpinES}).

After diagonalizing $H$ numerically, we estimated the density 
of states $D_0$ by averaging over all states with energies 
$E_n\in[-2\sigma_{\psi},2\sigma_{\psi}]$ 
(receiving about $95\,\%$ of the weight),
resulting in $D_0 \approx 962$.
Furthermore, we extracted $\bar\sigma_v^2$ [see below Eq.~\eqref{eq:ITR:AvgU4:f}] 
from the squared matrix elements $\lvert \bra{m} V \ket{n} \rvert^2$ within the 
relevant energy window in (\ref{x3}) (determined again by the $2\sigma_\psi$ 
criterion) by way of averaging 
around the diagonal within a band of $1000$ states,
yielding $\bar\sigma_v^2 \approx 0.0729$.
According to the definition below Eq. (\ref{eq:ITR:AvgU4:f}) 
this yields
\begin{eqnarray}
\alpha = \pi \bar\sigma_v^2 D_0 \simeq 220
\ .
\label{x4}
\end{eqnarray}
All parameters entering our analytical prediction~\eqref{eq:ITR:TypEcho} 
are thus explicitly available, i.e., there is no free fit parameter.

In Fig.~\ref{fig:ITR:SpinES}, we compare the numerical results obtained by 
exact diagonalization with our prediction~\eqref{eq:ITR:TypEcho} for different 
propagation times $\tau$ and perturbation strengths $\epsilon$, showing good agreement.
The largest deviations become apparent for small $\tau$ and large $\epsilon$.
By generalizing the analysis of Ref.~\cite{dabYY},
we expect that the band profile $\sigma_v^2(\Delta)$ of the perturbation 
$V$ becomes important in this regime.
The exponential form~\eqref{eq:ITR:TypEcho} for the suppression of the 
echo signal is then anticipated to show a transition to a Bessel-like decay 
of the form $4 J_1(x)^2/x^2$, where $x$ depends linearly on the noise 
strength $\epsilon$, the reversal time $\tau$, and the square root of the 
band width.
More generally, within the considered ensemble of inaccuracies the echo-signal attenuation is essentially given by the Fourier transform of $u(\gamma, E)$ introduced above Eq.~\eqref{eq:ITR:AvgU2},
which can be viewed as the ensemble-averaged fidelity or survival probability \cite{gor06, tor14} of an eigenstate $\ket{n}$ of the clean Hamiltonian $H$ under the imprecise backward Hamiltonian $H'$.
When the influence of the random inaccuracies in $H'$ increases, the echo signal may therefore be expected to approach the known decay of fidelity in pure random matrix models \cite{tor14}.
Unfortunately, we are not aware of an exact analytic solution of the pertinent equations in the intermediate regime between the exponential and Bessel-type behaviors \cite{dabYY}.

\section{Conclusions}
\label{sec:Conclusions}

We investigated the stability of observable
echo signals in many-body quantum 
systems under the influence of uncontrolled imperfections in the 
pertinent effective time reversal.
The considered protocol starts from a nonequilibrium 
initial state and lets 
the system evolve under the time-independent Hamiltonian $H$ 
for some time $\tau$,
at which an (effective) time reversal is performed that directs 
the system back towards the initial nonequilibrium state.
By introducing small inaccuracies in the time-reversed 
Hamiltonian, we obtained a measure for the instability of 
nonequilibrium states and the irreversibility of the dynamics
in terms of directly observable quantities.

Our prediction for the relative echo signal
 under such a distorted
backward Hamiltonian, Eq.~\eqref{eq:ITR:TypEcho}, includes an 
exponential dependence on both the squared perturbation strength 
$\epsilon^2$ and the propagation time $t$.
In particular, the height of the echo peak at $t \approx \tau$ is 
thus expected to decay exponentially in $\tau$.
Systems with this property were labeled ``irreversible'' 
in Ref.~\cite{sch16}, as opposed to ``reversible'' ones where 
the decay is (at most) algebraic.
In this sense, one may interpret the present result as a prediction 
that many-body quantum systems are typically irreversible.
However, it should be pointed out that the functional dependence 
of the echo peak on $\tau$ rather appears to be a property of the 
inaccuracies $V$ than of the system itself, even though the structure of 
$V$ is (in a real system) of course influenced by the 
properties of the system.
In any case, this functional dependence could also be confirmed 
for an exemplary spin-$\frac{1}{2}$ XXX model, whose dynamics 
shows good agreement with our analytical prediction
without any remaining fit parameter.

The paradigmatic example of macroscopic echo experiments 
are spin echoes and NMR \cite{hah50},
where nuclear spins precess in a strong magnetic field at different 
frequencies due to local inhomogeneities, leading to dephasing 
of the initially aligned magnetic moments.
Applying a $\pi$ pulse at time $\tau$ reverses the relative 
orientation between the spins and the external field, and thus 
effectively changes the sign of the 
corresponding term in the pertinent model Hamiltonian.
However, interactions among the spins and with the environment 
are not reversed and amount to a ``perturbation'' that causes 
deviations of the echo signal at time $2 \tau$, which -- in line 
with our general echo analysis here -- typically decays exponentially with $\tau$.
It should be noted that the ``imperfections'' in this context 
are usually vital in applications like MRI, precisely because 
different imperfections lead to slightly different decay rates 
and therefore allow one
to distinguish different materials. 

Experimental implementations of echo protocols are also available 
in a variety of \emph{interacting} spin systems, 
see, e.g., Refs.~\cite{rhi70, rhi71, zha92, lev98, usa98}.
In these experiments, an effective sign flip of the dominant part 
of the Hamiltonian (including dipole-dipole or even quadrupole 
interactions) is achieved by means of an elaborate application of 
radio-frequency magnetic fields.
The prevailing inaccuracies leading to deviations from the 
perfectly reversed signal are again due to nonreversible correction
terms in the Hamiltonian as well as possible experimental 
imprecisions in carrying out the required protocol.
Their major contribution is thus expected to be of the type studied here, too.

Specifically, the experimental study~\cite{lev98} indeed
reports an exponential decay of the 
peak height with the reversal time $\tau$ in a polarization echo 
experiment involving the nuclear spins of a cymantrene polycrystalline sample.
In the same study, data obtained from a ferrocene sample suggest 
an approximately Gaussian-shaped dependence, see also Ref.~\cite{usa98}.
The authors explain this with the much larger relative strength 
of the nonreversible component in the Hamiltonian,
compatible with our prediction that a crossover from an exponential 
to a Bessel-like \footnote{
Within the experimental error bars, the Bessel-like shape 
$4 J_1(2x)^2/(2x)^2$ is indistinguishable from a Gaussian shape $\exp(-x^2)$.
Note that the two functions agree to third order in $x$ and 
the relative difference of the fourth-order Taylor coefficients is only $1/6$.
}
decay is expected as the relative strength of $V$ increases, 
see the discussion at the end of Sec.~\ref{sec:Example}.

From a conceptual point of view, the approach of 
our present work, 
where the Hamiltonians of the forward and backward phases 
differ slightly, assesses (via observable quantities)
the stability of many-body trajectories with respect to variations of the 
dynamical laws.
Therefore, it should be no surprise that deviations grow 
with the propagation time $\tau$, and the exponential 
dependence might have been anticipated from 
perturbation-theoretic considerations, even though the 
applicability of standard perturbation theory is rather 
limited for typical many-body systems with their extremely 
dense energy spectra.
In that sense, the present derivation by nonperturbative 
methods is reassuring  and also indicates how deviations 
from the exponential behavior will manifest 
themselves if the influence of imperfections increases.
For future work, it will be interesting to investigate a 
complementary approach that studies the sensitivity towards variations of the initial conditions in macroscopic quantum system.

%%%%%%%%%%%%%%%%%%%%%%%%%%%%%%%%%%%%%%%%%%%%%%%%%%%%%%
\begin{acknowledgments}
L.\ D.\ thanks Patrick Vorndamme for inspiring discussions.
This work was supported by the 
Deutsche Forschungsgemeinschaft (DFG)
within the Research Unit FOR 2692
under Grant No. 397303734
and by the Paderborn Center for Parallel Computing (PC$^2$)
within the Project HPC-PRF-UBI2.
\end{acknowledgments}
%%%%%%%%%%%%%%%%%%%%%%%%%%%%%%%%%%%%%%%%%%%%%%%%%%%%%%%%

\appendix

\section{Derivation of Eq.~\eqref{eq:AvgTimeEvo:Rest:Bound}}
\label{app:AvgTimeEvo:Rest:Bound}

Exploiting Eq.~\eqref{eq:ITR:AvgU2}
we rewrite $R(t)$ from 
\eqref{eq:ITR:AvgTimeEvo:Rest}
as 
\begin{eqnarray}
R(t) 
& = & 
\tilde R(2 \alpha \epsilon^2 t)
\ ,
\label{a1}
\\
\tilde R(t) 
& := & 
\sum_{m,n} \bra{m} \rhoTarget \ket{m} \bra{n} A \ket{n} 
\, h(m-n, t)
\ ,
\label{a3}
\\
h(n,t)
& := & 
\frac{e^{-t}}{\pi N_v}
\, 
\frac{1 - \cos(t n/N_v) - \frac{\sin(t n/N_v)}{n/N_v}}{1+(n/N_v)^2}
\label{a4}
\ ,
\end{eqnarray}
where $N_v$ is defined in (\ref{eq:AvgU2:FWHM}),
and where $\omega_{kl} := E_k - E_l$
(see below Eq. (\ref{x2})) has been
approximated by $(k-l)/D_0$, 
as justified below Eq. (\ref{x3}).

According to~(\ref{eq:NumMixedLevels}),
we can and will take for granted that
\begin{eqnarray}
N_v\gg 1 
\ .
\label{a5a}
\end{eqnarray}
In view of (\ref{a4}) it follows that
the sum $\sum_n h(n, t)$ can be very well
approximated by the integral $\int \d x \,h(x, t)$.
Moreover, a straightforward but somewhat tedious
calculation yields $\int \d x \,h(x, t)=0$.
Therefore, we can subtract a constant value from 
the observable $A$ without changing the value of 
$\tilde R(t)$ in (\ref{a3}).
By means of the definition 
\begin{eqnarray}
\tilde A_{nn} := \bra{n} A \ket{n} - \< A \>_{\!\rhoMC}
\label{a6}
\end{eqnarray}
we thus can rewrite (\ref{a3}) as
\begin{eqnarray}
\tilde R(t) 
& = & 
\sum_{m} \bra{m} \rhoTarget \ket{m}\, \tilde Q(m,t)
\ ,
\label{a7}
\\
\tilde Q(m,t)
& := &
\sum_n \tilde A_{nn} \, h(m-n, t)
\ .
\label{a8}
\end{eqnarray}
Exploiting the Cauchy-Schwarz inequality yields
\begin{eqnarray}
\tilde Q^2(m,t)
\leq
\left[ \sum_n (\tilde A_{nn})^2 \right]
\left[ \sum_n  h^2(m-n, t) \right]
\ .
\label{a9}
\end{eqnarray}
Since the last sum over $n$ is independent of $m$
it follows that
\begin{eqnarray}
|\tilde Q(m,t)|
& \leq &
\sqrt{Q(t)}
\ ,
\label{a10}
\\
Q(t)
& := &
q(t) \sum_n (\tilde A_{nn})^2
\ ,
\label{a11}
\\
q(t)
& := &
\sum_n  h^2(n,t)
\ .
\label{a12}
\end{eqnarray}
Accordingly, $\tilde R(t)$ from (\ref{a7}) can be upper bounded
as
\begin{eqnarray}
|\tilde R(t)| \leq 
\sqrt{Q(t)}
\sum_{m} \bra{m} \rhoTarget \ket{m}
=\sqrt{Q(t)}
\ ,
\label{a13}
\end{eqnarray}
where we exploited that $\rhoTarget$ is 
a positive semi-definite operator of
unit trace.

\begin{figure}
\includegraphics[scale=1]{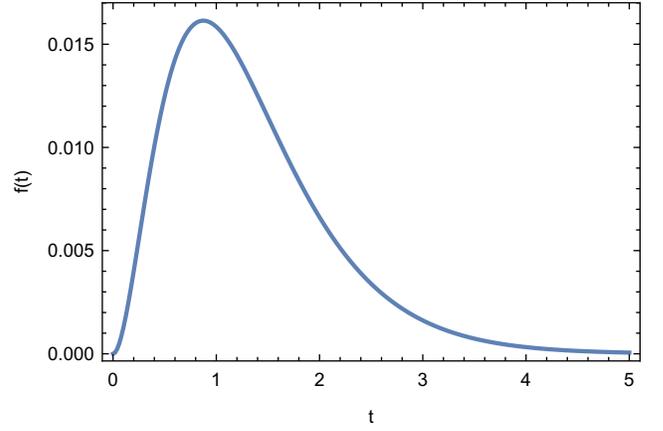}
\caption{
Numerical evaluation of the function
$f(t)$ from (\ref{a15}).
}
\label{fig2}
\end{figure}

Due to (\ref{a4}) and (\ref{a5a}) one
can conclude -- similarly as below (\ref{a5a}) -- 
that the sum on the right hand side of (\ref{a12})
is very well approximated by the integral 
$\int \d y \,h^2(y, t)$.
After going over from the integration variable
$y$ to $x:=y/N_v$ one thus obtains in very
good approximation
\begin{eqnarray}
q(t) 
& = &
N_v^{-1}\, f(t)
\ ,
\label{a14}
\\
f(t) 
& := & e^{-2t}
\int \d x \, \left[
\frac{1-\cos(xt)-\sin(xt)/x}{\pi [1+x^2]}
\right]^2
\, .\ \ \ \ \ \ 
\label{a15}
\end{eqnarray}

An analytical evaluation of $f(t)$ 
from (\ref{a15}) is possible but 
quite arduous, while a numerical
evaluation is straightforward,
see Fig. \ref{fig2}.
In either case, one finds that 
\begin{eqnarray}
0\leq f(t)\leq 1/50 =:c
\label{a16}
\end{eqnarray}
for all $t\geq 0$.
Taking into account (\ref{a1}), (\ref{a11}), (\ref{a13}),
and (\ref{a14}), we thus arrive at
\begin{eqnarray}
R^2(t) \leq \frac{c}{N_v}\, \sum_n (\tilde A_{nn})^2
\ .
\label{a17}
\end{eqnarray}

As discussed above Eq.~(\ref{x3}),
the diagonal matrix elements 
$\langle n|\rho(0)|n\rangle$
appearing in (\ref{eq:TimeEvo:A})
vanish whenever $E_n\not\in I_E$.
Accordingly, we can arbitrarily
modify the corresponding
$\langle n|A|n\rangle$'s
without any further consequences in 
(\ref{eq:TimeEvo:A}).
Specifically, we can 
%and will 
modify
them so that all $\tilde A_{nn}$
in (\ref{a6}) are zero if $E_n\not\in I_E$.
Therefore, the summation on the right
hand side of (\ref{a17}) can be
restricted to those $n$ with $E_n\in I_E$.
Altogether, we thus recover
(\ref{eq:AvgTimeEvo:Rest:Bound}) and
(\ref{eq:ETHViolation}) from the main text.

\end{document}